# Origin of the rich polymorphism of gold in penta-twinned nanoparticles

Camino Martín-Sánchez[a,b*], Ana Sánchez-Iglesias[c], José Antonio Barreda-Argüeso[b], Jean-Paul Itié[d], Paul Chauvigne[d], Luis M. Liz-Marzán[e,f], Fernando Rodríguez[b]

[a] Faculté des Sciences, Département de Chimie Physique, Université de Genève, 30 Quai Ernest-Ansermet, CH-1211 Genève, Switzerland
[b] MALTA Consolider, DCITIMAC, Facultad de Ciencias, University of Cantabria, Av. Los Castros 48, Santander, 39005, Spain
[c] Centro de Física de Materiales (CSIC-UPV/EHU), Paseo Manuel de Lardizabal 5, 20018 Donostia-San Sebastián, 20118, Spain
[d] Synchrotron SOLEIL, L'Orme des Merisiers St.Aubin, BP48, 91192 Gif-sur-Yvette, France
[e] CIC biomaGUNE, Basque Research and Technology Alliance (BRTA), Paseo de Miramón 194, Donostia-San Sebastián, 20014, Spain
[f] Ikerbasque, Basque Foundation for Science, Bilbao, 43018, Spain

Abstract

We report on the crystallographic structure of penta-twinned gold nanoparticles. Although gold typically exhibits a face-centered cubic (*fcc*) lattice, other phases have been reported in some nanoscale systems. We show that the crystallographic system and the lattice parameters of the gold unit cell strongly depend on the nanoparticle geometry, for a wide size range. Specifically, we show that decahedra exhibit a body-centered tetragonal structure (*I4/mmm*), whereas rods and bipyramids exhibit a body-centered orthorhombic structure (*Immm*). These changes in the crystallographic structure are explained by the elastic lattice distortions required to close the mismatch gap in penta-twinned nanoparticles, with respect to *fcc* single-crystal gold nanoparticles. The effects of nanoparticle shape and size on the surface pressure and the subsequent distortions are additionally discussed.

## Introduction

Noble metal nanoparticles (NPs) have attracted unprecedented interest in various fields, including e.g. sensing and biomedical applications [1-3], where the NPs' fascinating optical properties related to localized surface plasmon resonances (LSPRs) play a key role. Their large size-dependent extinction coefficients, with increasing ratios of scattering to absorption contributions, makes them attractive for use in thermal heating, evanescent field amplification, or highly sensitive plasmonic shifts in response to changes in the surrounding refractive index. These properties depend on the intrinsic optical properties of the metal - i.e., the plasmon resonances - and the NP shape and size [4-7]. In single-crystal nanoparticles – like in the bulk – metal atoms are ordered according to a face-centered cubic (*fcc*) structure, which is very stable even at pressures up to 200 GPa and temperatures up to the melting point [8,9]. The lattice parameters in nanostructured metals are slightly reduced under ambient conditions, compared to those in the bulk metal - typically about 0.1% for 10-100 nm in size - keeping the *fcc* structure stable for any size and shape within the pressure-temperature stability range of the single crystal nanoparticle [10-13].

At this point, the question arises whether this structure also applies to penta-twinned metallic nanoparticles (PT-NP). Previous works have revealed low-symmetry structural modifications from the *fcc* structure in penta-twinned gold microcrystals [14] and penta-twinned silver nanowires and nanodecahedra [15], towards tetragonal and orthorhombic structures. However, the true nature of the origin of these distorted structures still requires clarification because the reported results were obtained in microcrystals showing inhomogeneous strains [14] and in mixtures of different nanoparticle shapes in the studied colloids, yielding different crystal structures [15]. Furthermore, by means of X-ray diffraction (XRD) pattern analysis it has been shown that PT-Ag nanowires present a tetragonal structure plus some *fcc* components, with the cell volume of the tetragonal lattice being higher than that of the bulk Ag. By contrast, PT-Ag decahedra showed a tetragonal structure, albeit with some lattice parameters missing in the analysis.

Symmetry arguments reveal that PT-NP should indeed not be strictly cubic. In fact, the five (111)-type faces developing around the <110> twin axis cannot form a closed habit, as there is a mismatch angle of 7.5º between the five adjacent *fcc* single-crystal (tetrahedral) domains forming the PT-NP [16,17]. Either the creation of defects (stacking faults and dislocations) and/or slight modifications in the *fcc* structure can mitigate this structural gap. In this way, whether PT-NP with different geometries have purely distorted cubic structures or whether their geometry influences the crystal structure are open questions that deserve clarification. It is particularly intriguing, whether the surface pressure generated at the nanoscale can be sufficient to close the gap by elastically distorting the *fcc* structure of gold in PT-NP.

We present herein an XRD study on a set of highly monodispersed penta-twinned gold nanoparticle (PT-AuNP) colloids with different morphologies: nanorods (AuRod), bipyramids (AuBip) and decahedra (AuDec) (with two different sizes). We thus aimed to elucidate the crystal structure of PT-AuNP and to investigate whether it depends on the shape and size of the NP and whether any observed polymorphism can be explained by distortions of the *fcc* structure within the elastic theory.

## Experimental

**Nanoparticle synthesis:**

Penta-twinned AuNP with three different geometries - rod, decahedra, bipyramids - were synthesized *via* well-established seeded-growth methods, followed by surface functionalization with thiolated poly(ethylenglycol) which allows to have the gold nanoparticles colloidally stable in alcoholic mixtures [18]. The gold molar concentrations were set to achieve an optical density value around 70, so we could obtain suitable XRD diagrams for structural analysis [13]. Figure 1 shows representative TEM images and extinction spectra of the investigated PT-AuNP. AuBip had a mean length of 71±3 nm and a mean diameter of 19±1 nm, AuRod had a mean length of 55±2 nm and a mean diameter of 24±1 nm, and AuDec had mean side lengths of 31±1 nm and 49±1 nm.

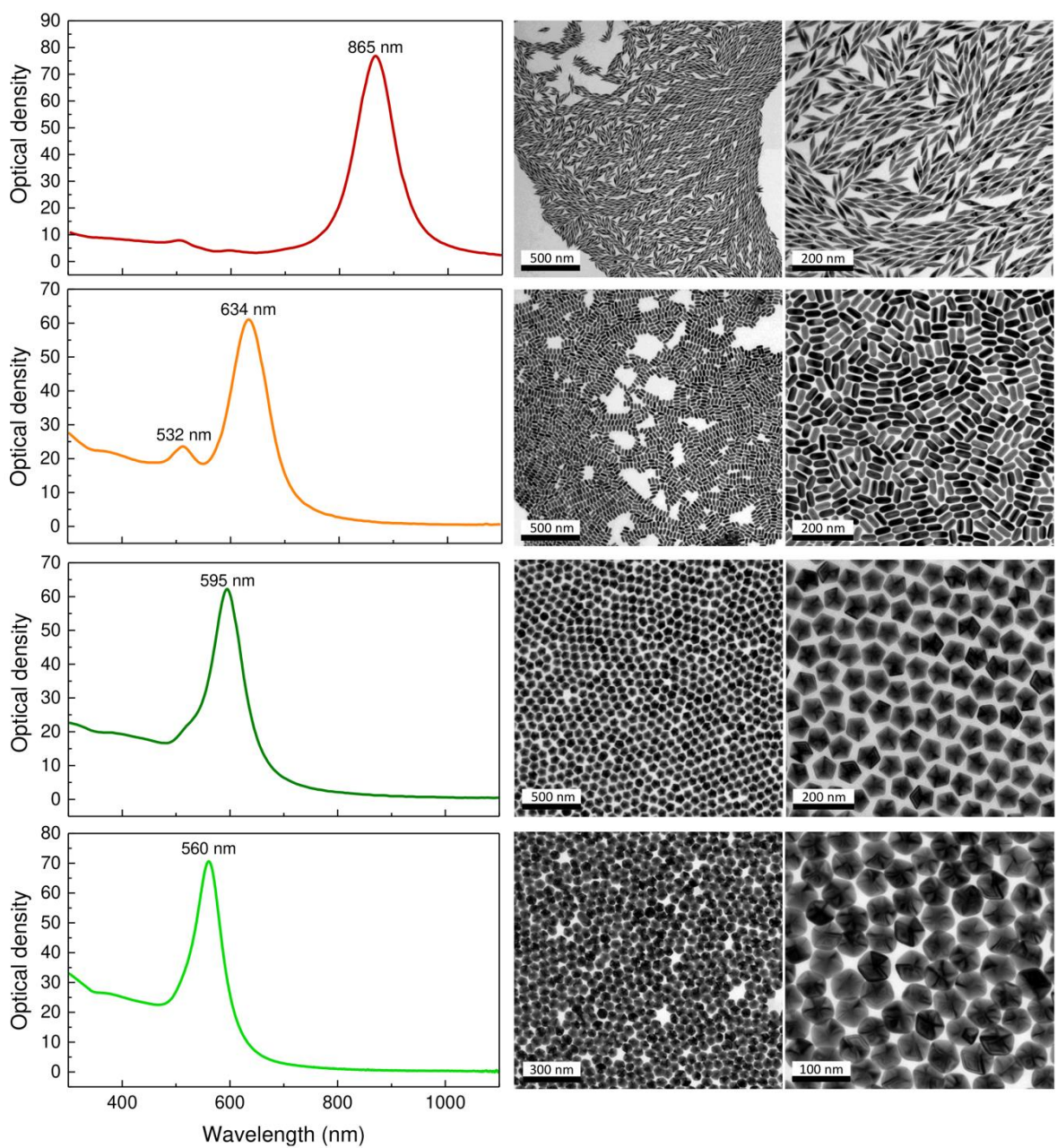

Figure 1. Optical extinction spectra and representative TEM images at different magnifications of the penta-twinned gold nanoparticles used in the experiments, from top to bottom: 71 nm×19 nm AuBip; 55 nm×24 nm AuRod; 49 nm AuDec; and 31 nm AuDec.

**X-ray diffraction measurements:**

XRD measurements on PT-AuNP MeOH-EtOH 4:1 colloids were performed at the SOLEIL Synchrotron (France) using the PSICHÉ beamline. PT-AuNP colloidal dispersions were measured in a diamond anvil cell (DAC), loaded into a 150 μm diameter hole within a rhenium gasket preindented to a thickness of 35 μm. Compacted gold powder of 2 μm average grain size was also loaded to precisely compare lattice parameters between systems under the same experimental conditions. A parallel configuration geometry for diffraction (incident X-ray beam parallel to the DAC load axis) was used. 2D XRD data were collected on a CdTe2M Dectris detector using a monochromatic X-ray beam with a wavelength of 0.3738 Å, focused to a beam size of 12×14 μm$^2$ (FWHM). The 2D XRD patterns were treated with the Dioptas program [19], and the intensity $I(2\theta)$ patterns were analyzed using the Match! Software [20]. LeBail fits were accomplished using Gaussian line profiles using two body-centered cells with a tetragonal *I4/mmm* space group for AuDec and an orthorhombic *Immm* space group for AuBip and AuRod. Instrumental resolution parameters were determined using a sample of $CeO_2$ with high crystalline quality using the Caglioti equation [21]: U = -0.009421, V = -0.008602 y W = 0.002219, to account for the instrumental broadening: $B_{inst}^2 = W + V \tan\theta + U \tan^2\theta$. The lattice parameters were determined with an accuracy better than 0.003 Å, and the FWHM of the Bragg peaks, $B_r$, were determined with a precision of 0.001° using the equation $B_{exp}^2 = B_{inst}^2 + B_r^2$.

# Results

Figure 2 shows the room temperature XRD patterns of all PT-AuNP colloids, as well as the corresponding bulk gold pattern for comparison. At first glance, all four XRD patterns resemble gold's *fcc* cubic structure (*Fm3m*). However, the XRD pattern of bulk gold is the only one that can be described in terms of an *fcc* structure with a lattice parameter of *a* = 4.0787(2) Å. The Bragg peaks for PT-AuNP show a splitting that is consistent with a body-centered tetragonal (*bct*) *I4/mmm* phase in decahedra and with a body-centered orthorhombic (*bco*) *Immm* phase in rods and bipyramids. The *bct* or *bco* cells can be seen as a reduction of the cubic cell, with the *a* and *b* parameters parallel to the cubic <1-10> and <110> directions, respectively, and the *c* parameter parallel to the <001> direction. In terms of the cubic cell, the tetragonal or orthorhombic parameters can be described as $a_{t,o} = \frac{a_c}{2}(\vec{i} - \vec{j})$, $a_{t,o} = \frac{a_c}{2}(\vec{i} + \vec{j})$, and $c_{t,o} = a_c \vec{k}$, with *a=b≠c* for the tetragonal system, and *a≠b≠c* for the orthorhombic system; the subscripts *c*, *t*, *o* refer to cubic, tetragonal and orthorhombic, respectively, and $\vec{i}, \vec{j}$, and $\vec{k}$ are the unitary vectors along the cubic cell. The low-symmetry cells have half the volume of the parent cubic cell. The measured lattice parameters of the PT-AuNP are *a=b*= 2.8990(7) Å and *c*=4.0320(11) Å for 49 nm decahedra, *a=b*= 2.8973(7) Å and *c*=4.0321(11) Å for 31 nm decahedra; *a*= 2.9150(7) Å, *b*= 2.8850(7) Å and *c*= 4.0208(11) Å for rods, *a*= 2.8986(7) Å, *b*= 2.8906(7) Å and *c*= 4.0401(11) Å for bipyramids. Interestingly, the unit cell volume of PT-AuNP decreases in all cases with respect to that of gold bulk, similar to what has been reported for single-crystal gold nanospheres and nanorods, as a consequence of surface pressure in nanoparticles [13]. In fact, the specific volume reduction - density increase - in PT-AuNP varies according to the nanoparticle geometry as 0.12% and 0.23% for 49 and 31nm AuDec, respectively, 0.33% for AuRod, and 0.22% for AuBip, which compares rather consistently with the 0.3% reduction found in single crystal AuNP of the same size [13]. It is worth noting that AuBip are synthesized in the presence of $Ag^+$ to induce their anisotropic growth. Inductively coupled plasma mass spectrometry (ICP-MS) measurements indicate that the content of silver in these AuBip is 3% [22]. Although silver has a larger *fcc* lattice parameter than *fcc* gold (4.0833 Å vs. 4.0787 Å) [23], the presence of silver atoms in the gold lattice (< 50 %) produces a reduction of the gold lattice [24,25]. Specifically, for a 3% silver concentration, the gold lattice parameter decreases by approximately 0.01%. After accounting for silver's contribution, the AuBip volume reduction is found to be 0.19% smaller than bulk gold Notably, the lattice volume of all PT-AuNP consistently decreases compared to bulk gold. Furthermore, for decahedral nanoparticles, the smaller size correlates with a more significant reduction in lattice cell volume.

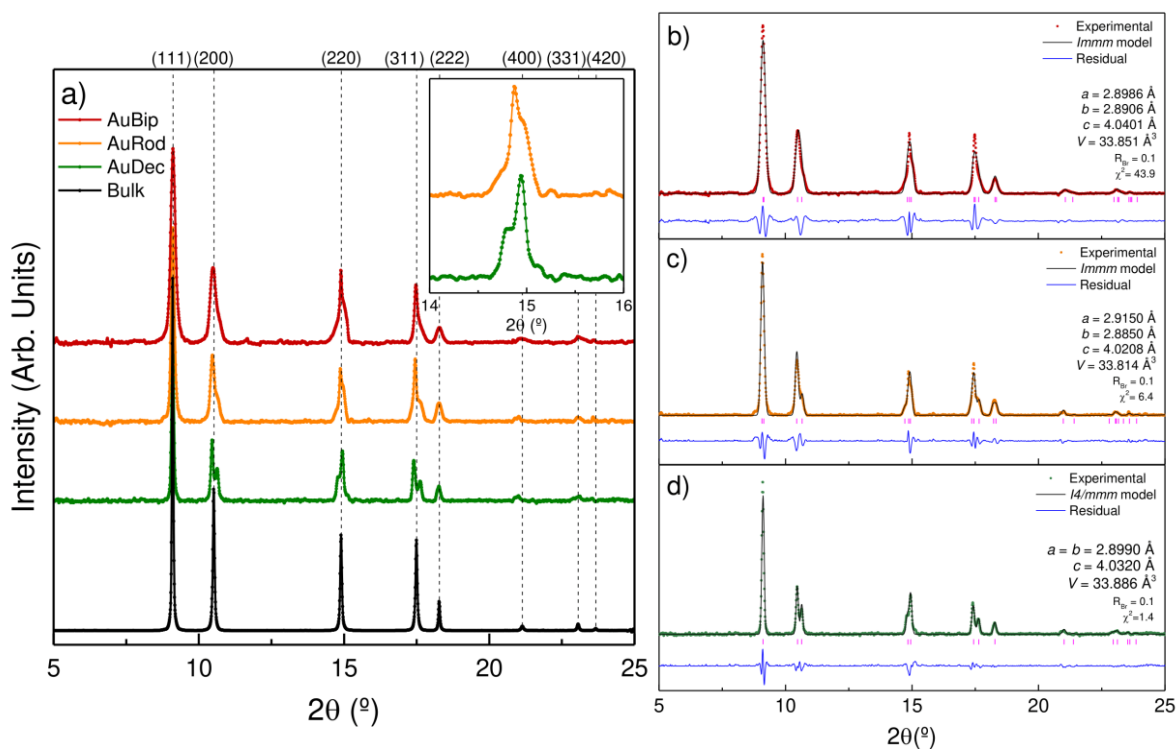

Figure 2. a) Diffraction patterns for AuBip (red), AuRod (orange), and AuDec (green) colloids in MeOH-EtOH 4:1 and Au micrometric powder (black). The inset shows a magnification of the cubic-equivalent (220) reflection of AuRod and AuDec. Note the different splitting of the reflection in the different geometries. Intensities were normalized to the (111) reflection. b,c,d) XRD patterns for AuBip, AuRod, and AuDec, respectively. Filled symbols correspond to experimental data; solid black lines correspond to calculated XRD model, and blue solid lines represent the fit residuum. The three PT-AuNP geometries reduce the specific volume with respect to the *fcc* bulk ($a_{fcc}$ = 4.0787(2) Å) by 0.1-0.2% for AuDec, 0.3% for AuRod, and 0.2% for AuBip.

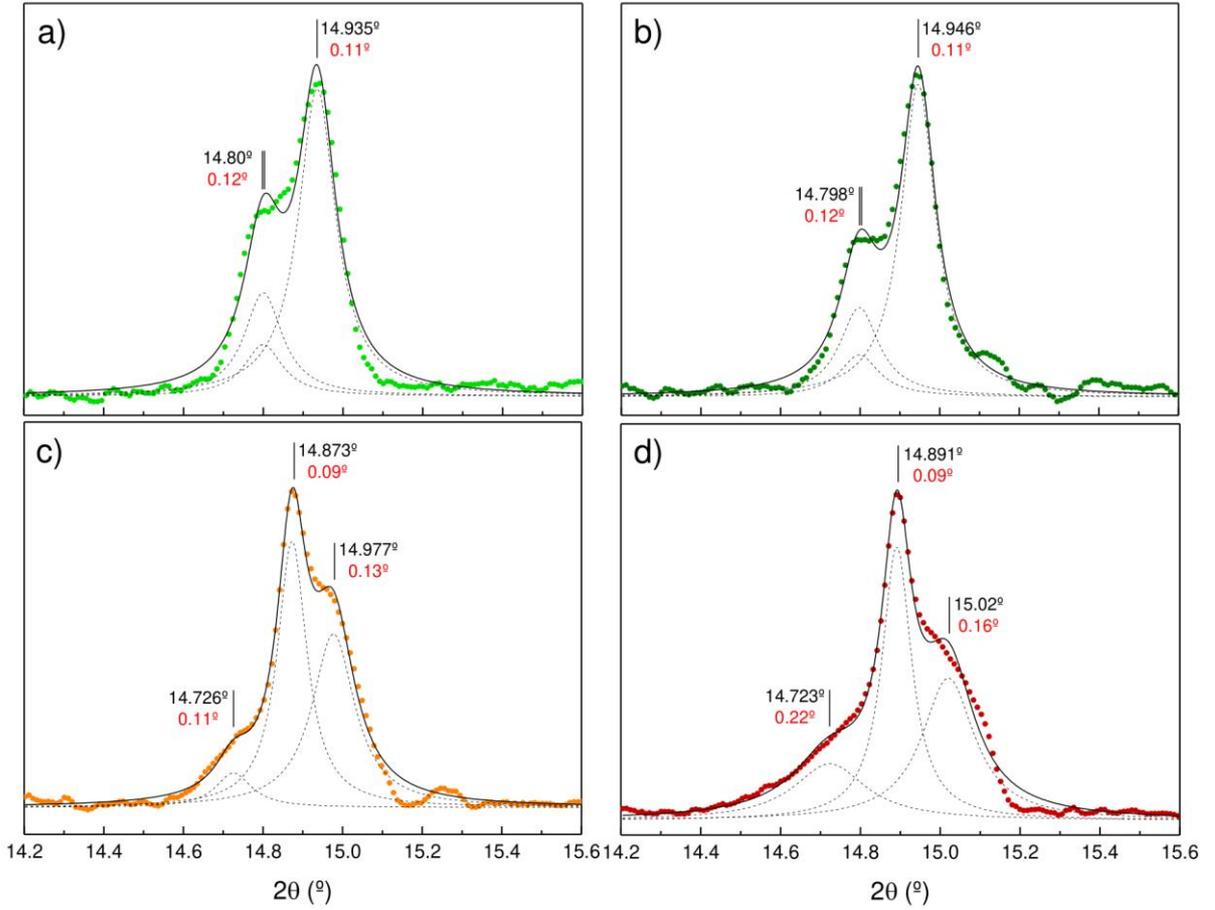

Figure 3. Cubic-equivalent (220) Bragg peak for a) 31 nm AuDec, b) 49 nm AuDec, c) AuRod, and d) AuBip colloids in MeOH-EtOH 4:1. Filled symbols correspond to experimental data; solid lines correspond to Lorentzian profile fitting to data; dashed lines indicate the deconvoluted peaks. Peak position and full width at half maximum of each contribution are indicated in black and red characters, respectively. Note the same tetragonal splitting pattern for AuDec, in contrast to the orthorhombic splitting pattern in AuRod and AuBip.

Although most of the observed (*hkl*) Bragg peaks resemble a tetragonal XRD pattern, the cubic (220) Bragg peak is the most sensitive one to distortions of tetragonal or orthorhombic symmetry and thus allows us to clearly distinguish between the two structures (see Fig. 3). Furthermore, the splitting of this peak contains direct information about the distortion of the cubic lattice in the final structure. The cubic (220) peak splits into two peaks in the tetragonal structure, whereas it splits into three components in the slightly distorted orthorhombic structure. As the orthorhombic distortion is weak, most Bragg peaks except (220) seem to fit into a tetragonal lattice. However, the Bragg peak of the (220) cubic plane family is, among the observed peaks, the only one that reveals the true symmetry of the lattice. Figure 3 shows how the (220) cubic Bragg peak splits into two peaks in AuDec, whereas it splits into three peaks in AuBip and AuRod. In the new *bco* orthorhombic space group, the Bragg angles $\theta_1$, $\theta_2$, and $\theta_3$ are associated with Brag peaks (200), (002), and (211), respectively, in order of increasing angle. Interestingly, lattice distortions can be determined directly from the angular positions $\theta_1$, $\theta_2$, and $\theta_3$. According to Bragg's law, parameters *a*, *b,* and *c* can be directly determined using the expressions $\sin\theta_1 = \sin\theta_{200} = \frac{\lambda}{a_{200}}$, $\sin\theta_2 = \sin\theta_{020} = \frac{\lambda}{b_{020}}$, and $\sin\theta_3 = \sin\theta_{112} = \sqrt{(\frac{\sin^2\theta_1}{4} + \frac{\sin^2\theta_2}{4} + \frac{\lambda^2}{c_{112}^2})}$. Additionally, lattice distortions $\frac{\delta l}{l}$ along the three orthogonal directions <1-10>, <110> –the twin axis–, and <001> are:

$$\varepsilon_{1-10} = \frac{\sqrt{2}a_{200}}{(2a_{200}b_{020}c_{112})^{1/3}} - 1 \quad (1)$$

$$\varepsilon_{110} = \frac{\sqrt{2}b_{020}}{(2a_{200}b_{020}c_{112})^{1/3}} - 1 \quad (2)$$

$$\varepsilon_{001} = \frac{c_{112}}{(2a_{200}b_{200}c_{112})^{1/3}} - 1 \quad (3)$$

and $a_c = (2a_{200}b_{020}c_{112})^{1/3}$ is the average cubic lattice parameter in the distorted structure, obtained as the cubic root of twice the *bco* cell volume. Table 1 collects the distortion parameters obtained by this method, and also summarizes the stresses associated with these strains in each PT-AuNP.

Table 1. Strains along the <1-10>, <110>, and <001> directions, and associated stresses, calculated using the elastic compliances of bulk gold [26,27]. Note that a negative or positive stress means compressive or tensile stress, respectively.

|  | $\varepsilon_{1-10}$ | $\varepsilon_{110}$ | $\varepsilon_{001}$ | $\sigma_{1-10}$ (GPa) | $\sigma_{110}$ (GPa) | $\sigma_{001}$ (GPa) |
|---|---|---|---|---|---|---|
| 31 nm AuDec | 0.0064 | 0.0064 | -0.0127 | 0.22 | 0.22 | -0.31 |
| 49 nm AuDec | 0.0066 | 0.0066 | -0.0130 | 0.20 | 0.20 | -0.32 |
| AuRod | 0.0127 | 0.0043 | -0.0158 | 0.76 | 0.11 | -0.22 |
| AuBip | 0.0119 | 0.0048 | -0.0165 | 0.54 | 0.00 | -0.40 |

Besides, the recorded XRD patterns provide interesting structural information. The FWHM of the Bragg peaks increases progressively, from AuBulk to AuDec 49 and 31nm (see Fig. 4), to AuRod and AuBip, proportionally to the reciprocal of the NP dimensions: 49 and 31 nm (AuDec), 24 nm (AuRod), 19 nm (AuBip). Interestingly, the $(220)_c$ cubic reflection splits into three components in AuBip and AuRod, showing different FWHM values. Reflection $(020)_o$, whose planes are perpendicular to the $<110>_o$ twin axis, is significantly narrower than reflections $(200)_o$ and $(211)_o$. This behavior is consistent with the higher coherence provided by planes perpendicular to the twin axis because the five reflections from the single-crystal domains forming the PT-AuNP diffract in the same direction, unlike $(020)_o$ and $(211)_o$ where each domain diffracts with the same Bragg angle but in a different direction, resulting in broader peaks.

Our XRD data indicate that the structure of PT-AuNP is not unique for a given crystallographic system; their polymorphism depends on the NP geometry. For AuDec, the structure is tetragonal regardless of the NP size. For elongated shapes like AuRod and AuBip, the structure is orthorhombic (see Figs. 2 and 3). To further explore this phenomenon, we developed a structural model within the elastic theory that aims to explain the different structures observed and provide rules to predict the structure of a PT-AuNP according to its geometry.

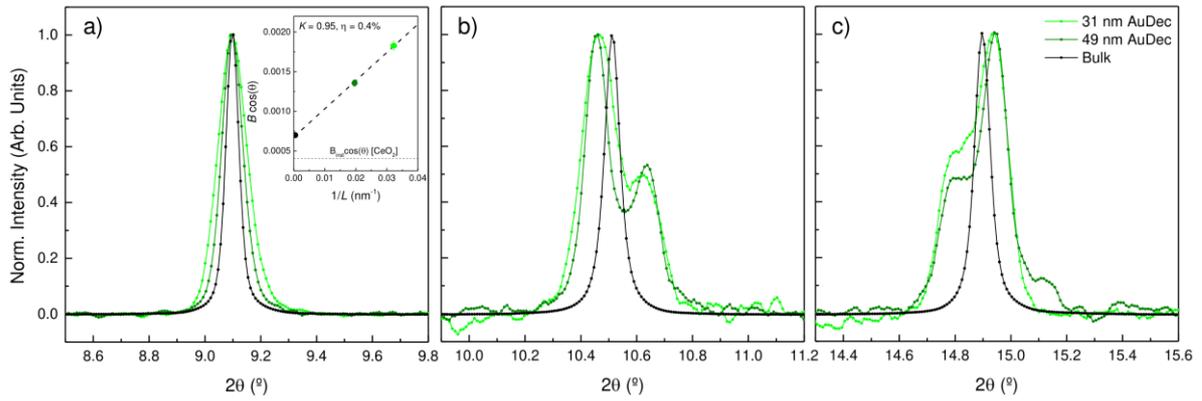

Figure 4. Cubic-equivalent a) (111), b) (200), and c) (200) Bragg peaks of 31 nm AuDec (light green), 49 nm AuDec (dark green) colloids in MeOH-EtOH 4:1, and Au micrometric powder (black). Note that cubic-equivalent (111) reflections have been shifted for width comparison purposes. The inset shows the Williamson-Hall plot illustrating the broadening of the cubic-equivalent (111) reflection with the decrease of the crystallite size.

**Elastic Model**

The distortions of the gold *fcc* cubic structure derived from XRD in PT-AuNP (see Table 1), for the three nanoparticle habits, indicate that there is a substantial reduction of the cubic lattice parameter along the $<001>_c$ direction –perpendicular to the $(001)_c$ planes– and an elongation along the cubic $<1\text{-}10>_c$ direction, i.e., a distortion perpendicular to the $<110>_c$ twin axis direction in the $(001)_c$ plane (see Fig. 5). These results suggest that the final structure of the PT-AuNP is determined by the lattice distortions required to close the geometrical gap of 7.5° in the penta-twinned nanoparticle, whose habit is characterized by a twin axis along the $<110>_c$ direction and five $(111)_c$ twin planes developing around it.

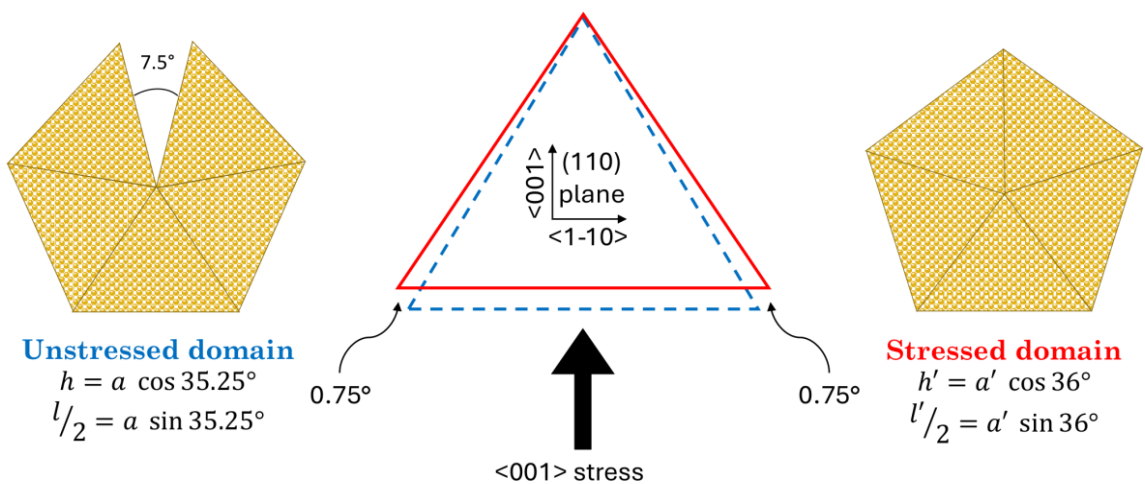

Figure 5. Schematic representation of the strains along the $<001>_c$ and $<1\text{-}10>_c$ directions – perpendicular to the $<110>_c$ twin axis- in each single-crystal domain forming the penta-twinned nanoparticle. $h$ ($h'$), $l$ ($l'$) and $a$ ($a'$) represent the height, base and edge of the triangles, respectively. The angle of 0.75° represents half the gap between neighboring *fcc* domains within the PT-NP. The blue dashed triangle ($h \times l \times a$) indicates the unstrained *fcc* domain, while the red solid triangle ($h' \times l' \times a'$) represents the elastically strained domain.

The model calculates the strain along the three orthogonal cubic directions <110>, <1-10>, and <001> needed to fill the gap, assuming that the gap can be partially filled up with staking faults or dislocations [17]. First, we calculate the strains within the (110) plane, denoted as $\varepsilon_{1-10}$ and $\varepsilon_{001}$, in each single-crystal domain of the nanoparticle:

$$\varepsilon_{1-10} = [1 + \cos^2 35.25\, \varepsilon_{001} + \sin^2 35.25\, \varepsilon_{1-10}] \frac{\sin \alpha}{\sin 35.25} - 1 \quad (4)$$

$$\varepsilon_{001} = [1 + \cos^2 35.25\, \varepsilon_{001} + \sin^2 35.25\, \varepsilon_{1-10}] \frac{\cos \alpha}{\cos 35.25} - 1 \quad (5)$$

Here, 35.25° is half the angle 70.5° of a single-crystal domain in a perfect cubic structure, and α is half the closure angle in the domain yielding distortions $\varepsilon_{1-10}$ and $\varepsilon_{001}$. This angle should be 36° if the gap closure can be entirely attributed to elastic deformations. Equations (4) and (5) give a relationship between the two strains, which, in principle, has multiple solutions for any arbitrary choice of one of the strains for a given value of α. To assess to what extent this purely elastic model explains the measured distortions, we calculated $\varepsilon_{001}$ as a function of the observed $\varepsilon_{1-10}$ distortion, using equations (4) and (5). The results are listed in Table 2. We find an exact match between calculated and observed distortions for a particular value of α, depending on the nanoparticle geometry. We obtain values of 35.8° for decahedra and 36.0° for bipyramids and rods, for the distortions required to fill the gap. The elastic strain in decahedra does not precisely match 36°, which we attribute to the partial filling of the gap by stacking faults and dislocations, as indicated elsewhere [17]. Besides, the model accounts reasonably well for the measured strains in the (110) plane (see Table 1). Decahedra are more affected by this extra gold filling, suggesting that the different shape of decahedra compared to the elongated bipyramids and rods plays a critical role. This different surface shape affects the surface stress acting on the PT NP. The more elongated the shape, the higher the stress will be on the lateral surface with respect to the axial stress. Energetically, decahedra can incorporate more gold atoms than rods and bipyramids to reduce the elastic energy due to the more homogeneous distribution of surface stresses in AuDec than in AuRod and AuBip. As we show below, this simple argument can explain why α is slightly smaller in decahedra than in rods and bipyramids.

Table 2. Experimental and calculated strains along <001> and <110> directions and associated closure angle of each structure. The experimental strain along <1-10> is taken as input parameter to calculate the strains along <001> and <110> from Eqs. 4 or 5, and 9, respectively.

|  | α(°) | $\varepsilon_{1-10,exp}$ | $\varepsilon_{001,exp}$ | $\varepsilon_{001,calc}$ | $\varepsilon_{110,exp}$ | $\varepsilon_{110,calc}$ |
|---|---|---|---|---|---|---|
| 31 nm AuDec | 35.8 | 0.0064 | -0.0127 | -0.0130 | 0.00640 | 0.00565 |
| 49 nm AuDec | 35.8 | 0.0066 | -0.0130 | -0.0129 | 0.00655 | 0.00503 |
| AuRod | 36.0 | 0.0127 | -0.0158 | -0.0156 | 0.00433 | 0.00372 |
| AuBip | 36.0 | 0.0119 | -0.0165 | -0.0163 | 0.00480 | 0.00476 |

Despite the strains in the (110) plane, the elastic strain along the twin axis <110> is irrelevant regarding the gap closure. However, it can be estimated by assuming that $\varepsilon_{110}$ provides the minimum elastic energy to the system, $\frac{\partial E_{elastic}}{\partial \varepsilon_{110}} = 0$. The elastic energy per volume unit of a structurally distorted cubic crystal can be written in terms of the strain-stress elastic compliances as:

$$E_{elastic} = \sigma_{110}\varepsilon_{110} + \sigma_{1-10}\varepsilon_{1-10} + \sigma_{001}\varepsilon_{001} \quad (6)$$

where the stresses correlate with the strains through the elastic compliances. The matrix related to the orthogonal coordinate set <1-10>, <110> and <001> is given by (in GPa units):

$$\begin{bmatrix} \sigma_{(110)} \\ \sigma_{(1-10)} \\ \sigma_{(001)} \\ \sigma_{(1-10)(001)} \\ \sigma_{(110)(001)} \\ \sigma_{(110)(0-10)} \end{bmatrix} = \begin{bmatrix} 213.2 & 136.3 & 160.5 & & & \\ 136.3 & 213.2 & 160.5 & & 0 & \\ 161.2 & 161.2 & 187.5 & & & \\ & & & 0.89 & & \\ & 0 & & & 0.46 & \\ & & & & & 0.35 \end{bmatrix} \begin{bmatrix} \varepsilon_{(110)} \\ \varepsilon_{(1-10)} \\ \varepsilon_{(001)} \\ \varepsilon_{(1-10)(001)} \\ \varepsilon_{(110)(001)} \\ \varepsilon_{(110)(0-10)} \end{bmatrix} \quad (7)$$

where the matrix elements have been calculated from the equations given elsewhere [28], using elastic the cubic elastic constants of the bulk gold [27] $C_{11} = 192$ GPa; $C_{12}=163$ GPa; $C_{44}=42$ GPa, referring to the orthogonal x, y and z coordinate axes which are parallel to the lattice vectors of the cubic cell. Writing the stress as a function of the strains, the energy derivative with respect to the <110> strain is given by

$$\frac{\partial E_{elastic}}{\partial \varepsilon_{110}} = 0 = 426.4\varepsilon_{110} + 272.6\varepsilon_{1-10} + 321.7\varepsilon_{001} \quad (8)$$

Therefore, we obtain

$$\varepsilon_{110} = -0.64\varepsilon_{1-10} - 0.75\varepsilon_{001} \quad (9)$$

Table 2 compares the measured $\varepsilon_{110}$ strain and the one calculated from the measured $\varepsilon_{1-10}$ and $\varepsilon_{001}$ strain values, using Eq. (9). Firstly, the agreement between measured and calculated strains supports the elastic model. The derived stress values indicate that there is compressive stress along <001> and tensile stress along <1-10>, both of which contribute to gap closure. However, the stress along the twin axis, which arises to minimize the elastic energy, is weaker and can be positive, negative, or zero depending on the relative stress on the (110) plane (Table 1). Interestingly, the volume reduction found for all three NP geometries (about 0.1-0.3%) corresponds to an effective surface pressure of 0.2-0.6 GPa. This surface pressure generated by the NP size is within the same order of magnitude as the stress required to close the gap of PT-AuNP (see Table 1). It must be noted that the stress difference $\sigma_{1-10} - \sigma_{110}$ is zero for AuDec consistently with the tetragonal structure, whereas it is non-zero for AuRod and AuBip, given their orthorhombic structure. Notably, the major specific surface area perpendicular to the twin axis in AuDec with respect to AuBip and AuRod can explain the differences in $\sigma_{1-10} - \sigma_{110}$ found for each geometry. This different stress distribution is responsible for the gap being able to close completely in elongated geometries, i.e. AuRod and AuBip. On the contrary, in AuDec the final stress distribution, yielding $\sigma_{1-10} = \sigma_{110}$, is insufficient to close the gap completely.

With this model, we conclude that the polymorphism in PT-AuNP is not associated with a given well-defined structure but depends on the geometry of the nanoparticle through partial filling of the gap *vs.* elastic distortions of the lattice, its shape affecting the stress distribution produced by the NP surface pressure. Thus, PT-AuNP with an elongated shape (rods or bipyramids) will fill the gap mainly with elastic strains, resulting in orthorhombic structures. Those NP geometries with a higher specific surface area perpendicular to twin axis (decahedra) will produced more equally stresses along $\sigma_{1-10}$ and $\sigma_{110}$ in the NP yielding tetragonal structures ($\sigma_{1-10} = \sigma_{110}$).

# Conclusions

We have shown that PT-AuNP exhibit a rich polymorphism, which can be described by tetragonal structures for decahedra and orthorhombic structures and rods and bipyramids. We propose that there is not a single polymorphic structure, because the crystallographic system and the lattice parameters of the unit cell depend on the nanoparticle geometry. The final structure of PT-AuNP is primarily related to the elastic lattice distortions necessary to close the nanoparticle gap attained in the *fcc* cubic phase of gold. These distortions depend on the geometry of the nanoparticle through the stress produced by the NP surface area. NP with an elongated shape, such as rods and bipyramids, will reduce the gap mainly through elastic distortions. In contrast, those with a more spherical shape will close the gap slightly, reducing elastic distortions at the expense of filling the gap with additional material, due to the more homogeneous stress distribution of $\sigma_{1-10}$ and $\sigma_{110}$ as the associated strains are insufficient to completely fill the gap. The NP shape plays a crucial role because the generated surface pressure is sufficient or comparable to the stress required to close the gap through elastic distortions. The final distortions of the cubic lattice, i.e., the lattice parameters of the tetragonal or orthorhombic phase, can be estimated from elastic theory. This knowledge is crucial for predicting how the crystal structure of PT-AuNP will vary with shape and size and how this will affect stability against compression or other physical properties (plasmonics, sensing, stiffness, etc.).

The aspect ratio of the PT-AuNP, which plays a relevant role in their plasmonic behavior, increases by about 2% with respect to the unstrained nanoparticle in the three studied geometries. In addition, the stress acting on the nanoparticle to close the gap provides a better mechanical stability to the PT-AuNP by compressing the nanoparticle in the direction perpendicular to the twin axis. This nanoparticle stress field likely explains why penta-twinned nanoparticles show better mechanical and thermal stability than single-crystal nanorods with the elongation axis of the rod along the cubic <100> crystallographic direction.

# Acknowledgements

Financial support from Projects PID2021-127656NB-I00, PID2023-151281OB-I00, and MALTA-Consolider Team (RED2018-102612-T) from the State Research Agency of Spain, Ministry of Science and Innovation is acknowledged. C.M.-S. acknowledges funding from the Spanish Ministry of Universities and the European Union-NextGeneration EU through the Margarita Salas research grant (C21.I4.P1). We acknowledge SOLEIL for the provision of synchrotron radiation facilities, and we would like to thank the staff for assistance in using beamline PSICHÉ (proposal 20230273).

# Abbreviations

AuBip, gold nanobipyramids; AuDec, gold nanodecahedra; AuRod, gold nanorods; bct, body-centered tetragonal; bco, body-centered orthorhombic; DAC, diamond anvil cell; fcc, face-centered cubic; LSPR, localized surface plasmon resonance; NP, nanoparticle; PT-AuNP, penta-twinned gold nanoparticle; PT-NP, penta-twinned nanoparticle; TEM, transmission electron microscopy; XRD, X-ray diffraction.